# Physics-informed Inverse Design of Multi-bit Programmable Metasurfaces

Yucheng Xu,[1,2] Jia-Qi Yang,[3] Kebin Fan,[1,2] Sheng Wang,[1,2] Jingbo Wu,[1,2] Caihong Zhang,[1,2] De-Chuan Zhan,[3] Willie J. Padilla,[4] and Biaobing Jin,[1,2] Jian Chen,[1,2] Peiheng Wu[1,2]

[1]*Research Institute of Superconductor Electronics (RISE) & Key Laboratory of Optoelectronic Devices and Systems with Extreme Performances of MOE, School of Electronic Science and Engineering, Nanjing University, Nanjing 210023, China*

[2]*Purple Mountain Laboratories, Nanjing 211111, China*

[3]*State Key Laboratory for Novel Software Technology, Nanjing University, Nanjing 210023, China*

[4]*Department of Electrical and Computer Engineering, Duke University, Box 90291, Durham, NC 27708, USA*

# Abstract

Emerging reconfigurable metasurfaces offer various possibilities in programmatically manipulating electromagnetic waves across spatial, spectral, and temporal domains, showcasing great potential for enhancing terahertz applications. However, they are hindered by limited tunability, particularly evident in relatively small phase tuning over $270^o$, due to the design constraints with time-intensive forward design methodologies. Here, we demonstrate a multi-bit programmable metasurface capable of terahertz beam steering, facilitated by a developed physics-informed inverse design (PIID) approach. Through integrating a modified coupled mode theory (MCMT) into residual neural networks, our PIID algorithm not only significantly increases the design accuracy compared to conventional neural networks but also elucidates the intricate physical relations between the geometry and the modes. Without decreasing the reflection intensity, our method achieves the enhanced phase tuning as large as $300^o$. Additionally, we experimentally validate the inverse designed programmable beam steering metasurface, which is adaptable across 1-bit, 2-bit, and tri-state coding schemes, yielding a deflection angle up to $68^o$ and broadened steering coverage. Our demonstration provides a promising pathway for rapidly exploring advanced metasurface devices, with potentially great impact on communication and imaging technologies.

# INTRODUCTION

The comprehension of the intricate interaction between light and metasurfaces has unlocked a myriad of routes to reshape the wavefront of electromagnetic waves at a subwavelength scale. Beam steering techniques made available with metasurfaces have garnered considerable attention, owing to its great potential for various technological applications, including light detection and ranging (LiDAR)[1, 2], wireless communications[3, 4], imaging[5–7], and augmented reality[8–10].

Recently, advancements in reconfigurable intelligent surfaces (RISs) within the microwave and terahertz range have underscored the indispensable role of beam steering in optimizing and directing the propagation of waves for next-generation wireless communications [11–14]. However, effective design approaches used in one frequency range may not be applicable to another due to limitations imposed by the dielectric properties of the selected materials. In the microwave range, dynamic beam steering has been successfully showcased through the integration of lumped elements, such as PIN diodes, and varactors [15–17]. These components facilitate substantial capacitance modulation on metasurfaces and typically introduce a typical small material loss. Consequently, a notable phase change of up to $300^o$ can be achieved while maintaining high amplitude, allowing multi-bit encoding schemes for beam steering. However, this methodology is not viable in the terahertz range because of the diminished on/off ratio of the semiconductor switches. Efforts to achieve reconfigurable terahertz metasurfaces involve hybridizing metasurfaces with active media, such as liquid crystal (LC), phase change materials, graphene, and semiconductors [18–24]. While these approaches have shown great success, unavoidable dielectric and Ohmic losses maintain values to those of the scattering rates. This, in turn, leads to a decreasing reflection and a reduced phase change around the resonant frequency.[25] Therefore, the quest for achieving a substantial phase change while preserving the amplitude in the terahertz and higher frequency ranges remains a significant challenge. To date, most terahertz programmable metasurfaces using regular shapes of crosses, circles, and squares are restricted to 1-bit reconfigurability, specifically phase tuning of $180^o$[26–29].

Since the electromagnetic response of metasurfaces is mainly governed by the geometry, it is conceivable that one may obtain highly performing metasurfaces by optimizing the structures. One option resorts to sweeping the geometric parameters based on human intuition and expertise, but such methods are extremely time-consuming and difficult to optimize owing to the vast parameter space. Inverse design methods, which rely on advanced computational algorithms to predict optimal designs given some prescribed electromagnetic properties, have been recently demonstrated and are a powerful strategy pushing the boundaries of human intuition[30, 31]. With the recent rapid development of artificial intelligence, these data-driven deep learning algorithms have brought new vitality to the inverse design of metasurfaces[32–38]. Feedforward deep neural networks (DNNs) trained with large amounts of data can accurately approximate the mapping from geometry to scatterings, such that it is possible to form a device which achieves the desired electromagnetic response[39, 40]. Deep learning-based inverse design, such as tandem neural networks[41], generative adversarial networks (GANs)[32, 43], variational autoencoders (VAEs)[43], adversarial autoencoders (AAEs)[44], neural adjoint (NA) method[45], and invertible neural networks (INNs)[46], have been successfully adopted for metasurface design and hasve demonstrated excellent performance. However, DNNs are regarded as 'black box' models ignoring the natural laws that dictate the electromagnetic response of metasurfaces. Therefore, as the DNNs become larger – requiring more training data - the interpretability and explainability become more challenging.

Research suggests that integrating prior physical knowledge into DNNs can mitigate their training burden and overcome limitations of traditional DNNs, especially when there are explicit governing equations for the problems [47–51]. In metasurface inverse designs, the main methods permitting DNNs to leverage physical knowledge include changing the model architecture, introducing explicit physical equations into loss functions, and modifying model parameters[52–56]. However, most of these studies rely on having an exact spectrum at the outset, which is often impractical in real-world applications.

Here, we propose a novel physics-informed inverse design (PIID) method for the design and realization of a high-performance programmable beam steering metasurface (PBSM). Our design approach leverages domain knowledge in the form of analytical equations developed based on a modified coupled mode theory (MCMT). Our MCMT incorporates etalon effects due to the superstrate, and this is integrated into a residual neural network as a physical layer. A key advantage of our approach is that it requires minimal information of the objective. Unlike other approaches, our algorithm allows inverse design only from three parameters: the desired phase tuning $\Delta\varphi$, reflection amplitude $R$ and operating frequency $f_t$. The additional physics layer also allows us to uncover the relationship between the geometries and the metasurface resonance properties. Using this method, we perform the inverse design of a metasurface which exhibits a phase tuning of approximately $300^o$ with a reflection amplitude over 90% at 465 GHz. Finally, we experimentally validate our approach through the creation of a multi-bit programmable metasurface. The metasurface can be coded in 1-bit, 2-bit, and tri-state sequences arbitrarily, and is therefore able to steer a normal incidence beam up to deflected angles $68^o$. Our work verifies that the incorporation of physics knowledge can significantly simplify the inverse design process for high-performance metasurfaces, which are crucial for various applications.

## RESULTS

**Working Principles:** Generally, achieving large phase tuning $\Delta\varphi$ over 270 degrees and high reflection $R_t$ at the working frequency of $\omega_t$ are crucial for designing high-efficiency reflective beam-steering metasurfaces and holograms with multi-bit coding modality [36,57]. According to temporal coupled theory, these conditions necessitate the metasurface to remain in an overcoupled state, characterized by a large radiative loss rate and a significant resonant frequency shift. ((see

Sec. 3 in Supplementary information for detailed analysis) In the microwave, these requirements are easily met due to negligible material losses. However, in the terahertz range and beyond, where dielectric and Ohmic losses become substantial, designing dynamic metasurfaces with such extensive phase tuning becomes remarkably challenging. It is also not easy to link the simulated reflection spectra to the physical parameters of resonators-including material and radiative loss rates-using traditional forward-optimization methods through sweeping geometries. Furthermore, the design target with three given parameters is not enough to design a dynamic metasurface with tens of geometric parameters since such a design is significantly ill-posed. Here, we proposed an inverse design strategy leveraging the efficacy of deep neural networks and the mode parameters of metasurfaces. Fig.1 shows a schematic diagram of a reconfigurable metasurface for beam steering designed from our proposed PIID. The design process starts from a given design target set, which includes three prerequisites for multi-bit programmable metasurfaces. In the next section, we develop a multilayer coupled mode theory to show that these three quantities are dominated by the resonant characteristics of metasurfaces $\Omega$, which includes the resonant frequency $\omega_i$, radiative loss rate $\gamma_i$, and material loss rate $\delta_i$ at $i$ = On/Off states, respectively. These parameters well describe the reflection spectra of metasurfaces. We next use residual MLP (ResMLP) neural networks which can establish a mapping between the metasurface geometry and these characteristic parameters. Such a mapping allows a fast inverse design of metasurfaces with desired performance in seconds. Then, the most suitable design is selected for constructing the programmable beam-steering device.

**Modified coupled-mode theory for metasurface on a superstrate:** As shown in Fig. 2a, the dynamic metasurface consists of four layers with LC sandwiched between the top metallic resonator and bottom ground plane. The top layer is a 300-$\mu$m thick quartz substrate to support the metallic resonators. All resonators in the same column are connected to form a linear array. To increase the geometric freedom of the resonator, a B-spline algorithm with a total of 13 control points ($r_1 \sim r_{13}$) is adopted to generate a freeform geometry in the first quadrant as shown in Fig. 2a. Then the pattern is generated through mirroring concerning the xz plane and yz plane, respectively. In addition to the 13 control points, periodicity $p$ and height of LC $d_s$ are also important geometric parameters to define the resonant properties. Finally, a total of 15 geometric parameters constitute the design space: $g = [p,d,r_1,r_2,...r_{13}]$, which are also used as the input during the training.

Generally, the fundamental resonance of a metasurface with a metal-insulator-metal (MIM) configuration behaves as a Lorentzian response, which can be well described using the coupled-mode theory with the important characteristic parameters $\Omega = \{\omega_0, \gamma, \delta\}$, which are resonant frequency, radiative loss rate and material loss rate, respectively. Therefore, the retrieval of the metasurface reflection spectrum around the resonance is possible as long as the three parameters are known. In our model, however, the extra 300-$\mu$m quartz substrate on top of the resonator leads to a Fabry-Pérot effect, thereby reshaping the Lorentzian response significantly, as the red solid line shown in Fig. 2c. To explicitly describe the complex spectra, we consider the MIM layer as an equivalent thin film layer coated on a superstrate, which is modeled in Fig. 2b. Then the total reflection can be developed using the transfer-matrix method[58]. The reflection of the equivalent metasurface layer is given by the coupled-mode theory as

$$r_i^m(\omega) = -1 + \frac{2\gamma_i}{j(\omega - \omega_{0i,l}) + (\delta_{i,l} + \gamma_{i,l})} + \frac{2\gamma_{i,h}}{j(\omega - \omega_{i,h}) + (\delta_{i,h} + \gamma_{i,h})} \quad (1)$$

where the subscript 'i=On/Off', corresponds to the on and off of the applied voltage on the LC; $l$ and $h$ denote the fundamental and higher-order resonances, respectively. Here, two resonances are considered simultaneously due to the broadly monitored frequency range. Then, the transfer-matrix method is employed to derive the total reflection from the composite structure, which is simplified (Sec. 4 in Supplementary information for derivation) as:

$$r_i(\omega) = \frac{-j + (\eta - j\xi)r_i^m(\omega)}{(\eta + j\xi) + jr_i^m(\omega)} \quad (2)$$

with

$$\eta = \frac{2n}{n^2 - 1}cot(nk_0 d_s), \quad \xi = \frac{n^2 + 1}{n^2 - 1}$$

where $n$ is the complex refractive index of quartz, $k_0$ is the wavenumber in free space and $d_s$ is the thickness of the quartz superstrate. For a given refractive index and thickness of the substrate, $\eta$ only varies with the frequency. The metasurface response $r_i^m(\omega)$ obtained using Eq. 2 is shown in Fig. S1. Then, the characteristic parameters can be fitted through the amplitude of Eq. 1. As fitted characteristic values are substituted into Eq. 2, the dotted lines shown in Fig. 2c indicate that the calculated reflection amplitudes of randomly generated LC metasurface agree with the simulations (solid lines) very well at On/Off states, respectively. Further, the good agreement on the phase marks the validity of our model since only reflection amplitude was applied for fitting. Based on the MCMT, two reflection spectra at On/Off states can be quickly calculated from two sets of $\Omega$.

**Physics-informed neural networks:** A multi-bit beam-steering PBSM normally necessitates the unit cell which achieves a large phase tuning range while keeping the amplitude as high as possible at the operating frequencies to ensure the high efficiency of the beam steering. Therefore, these requirements are equivalent to the final design targets shown as $\omega_t$, $R_t$, and $\Delta\varphi_t$. In principle, the design targets could be analytically determined by Eq. 1. However, for a general dynamic metasurface design with varied radiative loss and material loss rates, such an analytical method is challenging. As a result, we applied an adaptive gradient algorithm to rapidly calculate the mode parameters based on the required design targets. The objective function defined for phase difference is as follows:

$$L(\Delta\varphi) = [(\Delta\varphi - \Delta\varphi_t)]^2 \qquad (3)$$

Then, the phase difference with the resonant loss and the radiation loss rates are calculated based on $\frac{\partial L(\Delta\varphi)}{\partial \omega_{off}}, \frac{\partial L(\Delta\varphi)}{\partial \omega_{on}}, \frac{\partial L(\Delta\varphi)}{\partial \delta_{off}}, \frac{\partial L(\Delta\varphi)}{\partial \delta_{on}}, \frac{\partial L(\Delta\varphi)}{\partial \gamma_{off}}, \frac{\partial L(\Delta\varphi)}{\partial \gamma_{on}}$. The reflection coefficient $R_t$ at the target frequency $\omega_t$ will be evaluated after $\Delta\varphi_t$ is reached. To accelerate convergence, the initial frequency is set as the mid-point of the two modal frequencies, i.e. $(\omega_{on} + \omega_{off})/2$, since our analysis indicates that the frequency with maximum phase modulation is around the mid-point of the two resonances. Further, according to the statistical analysis, we also noticed that the frequency achieving maximum phase modulation is around the intersection of the two reflection spectra (see Sec. 6 in Supplementary information for the analysis). Therefore, the amplitude is only adopted as an additional criterion, and $\Delta\varphi_t$ should be viewed as the main design target.

After obtaining the target mode parameters, we proposed a Residual MLP (ResMLP) network architecture, to delineate the mapping from metasurface geometry to mode parameters. To simplify our neural network for physical modeling, we replaced the convolution layers, which are used in typical Resnets, with linear layers. Additionally, we augmented the architecture with batch normalization layers following the nonlinear layers to enhance convergence speed. It integrates three stacked Bottleneck Residual MLP Blocks (BRMB) as shown in Fig. 1a, with linear layers at the input and output ends orchestrating dimensionality mapping to suitable proportions. To enhance the network's convergence pace, a residual connection is introduced after the input layer to the second Linear layer. Employing a Bottleneck design between the two linear layers yields a tenfold reduction in computational load compared to using identical dimensions in a traditional MLP. The smaller intermediate layers aid in curtailing extraneous information, thereby elevating generalization performance. Different from conventional MLP, this structure boasts expedited processing and convergence, coupled with an augmentation in generalization capabilities (See Sec. 7 in Supplementary information for more details about the neural networks). In the forward model, 12000 designs including 'On-state' and 'Off-state' reflection spectra and retrieved corresponding mode parameters were used for training and another 3000 datasets were used for validation. In both processes, the calculated loss of ResMLP converges 2.5 times faster than a conventional MLP but with 10 times larger hyperparameters. After setting up the forward neural networks, the neural adjoint method was directly applied to inversely design the metasurface[46]. In addition to three design targets, to ensure the steering angle of the designed device and low voltage applied to modulate the metasurface, another two constraints on the periodicity and LC height were also added to the loss function of the inverse design. (See Sec. 7 in Supplementary information for more details) The inverse design procedure generates a total of 500 candidates in 10 seconds. Then all the candidates are fed into the forward model to verify the design, and 10 out of 500 will be selected for simulation verification and the best one will be used for further experimental demonstration.

Figure 3b shows the simulated reflection spectra at 'On' and 'Off' states, respectively, from the inverse designed LC metasurface with design targets as $\omega_t$=435 GHz, $\Delta\varphi_t$=300$^o$, and $R_t \geq 0.9$. The geometric constraints are set with the periodicity of 240 $\mu$m and LC height of 10 $\mu$m. (Table. S2 in supplementary material for geometric details) This example indicates the efficacy of designing a dynamic metasurface with phase tuning over 270$^o$ and reflectance over 0.9 in the terahertz range, which has not been demonstrated before.

The integration of physical layers into the neural networks can further enhance the interpretability of complex mapping due to the direct link between the geometry and physics. An example with a target operating frequency of 480 GHz is used for demonstrating the ability to learn underlying physics from the networks. Figure 3c shows the correlation strength of the predicted mode parameters and the input geometric parameters. Notably, the thickness of the LC layer $d$ and control parameter $r_7$ exert the most significant influence. The radiative loss and material loss rates are closely correlated to the LC thickness, while, surprisingly, the fundamental resonant frequency is mainly controlled by the size in the diagonal direction. To show the significance of $r_7$, we performed a numerical comparison via manually changing the $r_7$ and $r_{13}$ at the 'Off' state, respectively, as shown in Fig. 3d and e. Both the resonant frequency and bandwidth of the spectra exhibit larger sensitivity to $r_7$. Further analysis reveals an explicit relation between the geometric and the mode parameters, as shown in Fig. 3f. The predicted relations between the resonant frequencies at 'On' and 'Off' states and the periodicity match with numerically simulated relations very well. These relations are also very similar to those reported in other literature[59]. Figure 3g shows similarly excellent predictions of the correlation between the resonant frequency and LC thickness, manifesting the interpretability and remarkable prediction accuracy of the PIID. Overall, our proposed PIID outperforms that using a similar ResMLP without the physics layers significantly. Figure 3h illustrates much smaller statistical mean squared error (MSE) values calculated from 500 inverse designs generated by ResMLP embedded with physics layers than directly training on the spectrum without informed physics.

**Demonstration of Programmable Beam Steering Metasurface:** In this section, we will experimentally demonstrate a multi-bit programmable LC metasurface using PIID. To design a metasurface with two-bit programmable capability, the design targets are set as follows: $\omega_t$=462 GHz, $\Delta\varphi_t$=270º, and $R_t \geq 0.8$. Two geometries, *i.e.* the period and LC layer thickness, are fixed at 240 $\mu$m and 10 $\mu$m, respectively. The details of the geometry are listed in the Sec. 8 in Supplementary information. Fig.4a shows the schematic of the unit cell obtained through PIID. The simulated reflection of the designed structure exhibits a frequency redshift of 37.5 GHz from 481.8 GHz to 444.3 GHz due to the increased permittivity of LC after the voltage is applied between the top resonator layer and the ground plane. In addition, at the intersection of the two spectra at the 'On' and 'Off' states, the reflection amplitude is as high as 0.86, and the phase change at this frequency is about 268.2º, as shown in Fig. 4b. Both the amplitude and the phase change match the design targets very well. Then, a sample consisting of 112×112 units was fabricated with the microscopic image shown in Fig.4c and 4d. The dynamic tuning of the sample was verified using time-domain terahertz spectroscopy with all the units applied with the same voltage. (Sec. 2 in Supplementary information for the setup details) Under the normal incident, the measured Off-state resonant frequency is at 485.2 GHz as the blue curve shown in Fig. 4e. when 15 V is applied, the resonance redshifts to about 451.6 GHz with a total frequency shift of 33.6 GHz. The amplitude at the intersection of 467 GHz is about 0.7, which is a little lower than the design. The reflection dips at the resonance frequencies are also a little lower than the simulated results, which can be attributed to the metal loss induced by the sputtering and a slightly higher loss tangent of liquid crystal than that used in the modeling. Such an unexpected material loss also causes the phase jump of the sample at the "Off state", which is normally seen when the metasurface transits from an over-coupled condition ($\delta < \gamma$) to an under-coupled condition ($\delta > \gamma$) [60], as shown in Fig. 4f. Nevertheless, the phase tuning at the intersection frequency is measured to be 264.8º, which is close to our design target.

To demonstrate the ability to encode multi-bits for beam steering, the units in the same column are connected, so that a total of 56 columns can be independently controlled through an FPGA. To ensure an accurate gradient phase applied on the metasurface, the phase changes as a function of the electric bias with an interval of 1V are characterized as shown in Fig. 5a. To be consistent with the beam steering experiments, the characterization of phase change was implemented with an incidence tilted by 10.5º. As a result, the measured maximum phase change is 253.0° at 465.4 GHz, which is about 12° smaller than that characterized under normal incidence as shown in Fig.4f.

According to the generalized Snell's law[61], the deflected beam is controlled by the phase gradient at the interface of two media. For a metasurface consisting of supercells with periodicity comparable to the wavelength, non-local effects caused by the grating-like supercells will also contribute to the beam deflection [62, 63], which can be formulated as:

$$sin\theta_{defl} - sin\theta_{inc} = m\frac{\lambda}{S} + \frac{\lambda}{2\pi}\frac{d\varphi}{dx} = (m+1)\lambda \qquad (4)$$

where $S$ represents the periodicity of a supercell, $m$ is the diffraction order, and $\lambda$ is the operating wavelength. Through modifying the phase gradient in a supercell, the deflection angle will be tuned. Then, programmable beam steering was demonstrated using the designed metasurface with applying 1-bit, 2-bit, and tri-state coding sequences. The operating frequency for the 1-bit and 2-bit coding is at 462 GHz, as the red dashed line shown in Fig. 5a, while that for tri-state coding is at 457 GHz shown as the black-dashed line. In the experiment, the incidence was tilted by 10.5º from the normal direction. Figure 5d-e show the measured results for applying the 1-bit coding sequence to a supercell consisting of two columns with the applied voltage and phase shown in Fig. 5b. Finally, the maximum deflection angle can be as large as 68º at the frequency of 442 GHz, which is a little higher than the prediction of 62.8º as shown in Fig. 5e. The deviation of the operating frequency from the design could be caused by the modified phase on each column from mutual coupling. And the small reflection amplitude at 68º is mainly caused by the weak reflection with an 8V bias.

In the 2-bit beam steering scheme, four subcells with a phase difference about 90º will be utilized to form a supercell. Based on the experimental characterization, four subcells with applied voltages of 0 V, 7 V, 8 V and 15 V, corresponding to phase change of 0º, 74.3º, 177.2º, and 253.0º, are encoded as '00', '01', '10', '11', respectively, as shown in Fig. 5b. The measured angular deflection is about 31.6$^o$ at the frequency of 457 GHz with an amplitude of 0.26, which is much stronger than the 1-bit coding method. In addition to the 1-bit and 2-bit coding sequences, a tri-state coding sequence can also be applied to the PBSM, as shown in Fig. 5h-i. Such a coding sequence has not been experimentally demonstrated in the terahertz before. In our experiment, a supercell consisting of 6 columns was encoded as '0', '1', and '2' as shown in the inset of Fig. 5i. The subsequently applied voltages of '0V', '10V', and '15V' correspond to phases of '0$^o$', '120$^o$', and '240$^o$'. The measured beam at 457 GHz is deflected by 23 degrees, which agree with the calculation very well. It is noteworthy that the deflected angles by tri-state coding are not attainable by 1-bit and 2-bit coding sequences. In Fig. 5c, we further explore the deflection-angle coverage space of a 2D row-column controlled reconfigurable device based on our designed metasurface.[29] The coverage of the 2-bit coding scheme overlaps with the 1-bit coding scheme but with a higher deflection efficiency. In addition to coverage by 1-bit coding sequence (covering area is $\theta < 60^o$ and $0^o < \varphi < 360^o$), it also exhibits more coverage at an elevation angle of about $\theta = 25^o$, which significantly improves the angular resolution of the deflected beam. This increased coverage is mainly contributed from the tri-state coding scheme shown as the red dots in Fig. 5c.

## DISCUSSION

In agreement with the design targets, the reflectance at the operating frequency remains high when 0V and 15V are applied on the metasurface. However, the measured reflection is very low when the voltage is changed to 7V and 8V. We attribute this to the extra loss introduced into the metasurface such that it operates very close to the critical coupling condition, with nearly zero reflection at the resonant frequency. This problem can be alleviated via using a thicker LC layer since the learned physics from the neural networks shows that the loss rate decreases with the LC thickness while the radiative loss increases with the LC thickness. (See Sec.9 in Supplementary information) As a tradeoff, the thicker LC layer would also increase the bias voltage to ensure the large birefringence of the LC.

In addition to the single-frequency operation, the programmable metasurface is also possible to operate at multiple frequencies and broad frequency ranges. In this work, the metasurface was only designed at a single operational frequency, however, the coupled model can be well adapted to describe a multi-resonant system[64]. Therefore, by adding more mode parameters into the physics layer and increasing the complexity of the metasurface geometry, it could obtain a programmable metasurface with multiple deflection angles simultaneously only from a linear array. Further, as shown in Fig. 5a, the phase change induced by the 15V in the range of 460 GHz and 475 GHz does not vary significantly, indicating a broad operation range for beam steering as shown in Fig. S11 in Supplementary information. It is also noteworthy that the bandwidth of the phase change is much dependent on the resonant frequency shift. Then, the operational bandwidth for programmable metasurface could be further increased through designing metasurfaces with a large frequency shift, such as using LC with a larger birefringence and covering the metasurface layer with modified graphene layer to ensure all the LC mesogens to rotate under the graphene besides those under the metallic resonators[65].

In conclusion, we demonstrated the inverse design of a multi-bit terahertz PBSM using physics-informed ResMLP neural networks with only three target parameters. The physics layer in the network, developed based on a modified coupled mode theory, takes the superstrate effect into account, and not only enables a much faster learning of the network, but also further enhances the interpretability of the networks. The experiments on beam steering using the PBSM showcase a remarkably large steering range from 0 to over 68$^o$ using 1-bit, 2-bit, and tri-state coding schemes. We believe that our method paves a new way for efficiently customizing reconfigurable and programmable meta-devices for various applications, including 6G wireless communications, hyperspectral imaging, and holographic displaying.

## Methods

Fabrication: First, the metasurface layer was defined by photolithography on a 300-μm thick quartz superstrate. Then, a 300-nm gold film was deposited on the superstrate and a substrate quartz, which will be used as a ground plane. The metasurface layer was obtained by the lift-off process. Next, both of the two quartz slabs were coated with the aligning agent of NC-M-SD1 and exposed to ultraviolet light with an average energy of 3 J/cm2 for 10 minutes to complete the initial orientation of LC. After that, UV glue mixed with silicon spheres with an average diameter of 10.05±0.05um was coated on the edge of the substrate via a dispensing machine. After 30 seconds solidification of the UV glue, the LC was heated to 118°C and infiltrated into the gap. Then, the LC was sealed into the chamber using the UV glue. Finally, the sample was mounted on a PCB board and wire bonding was implemented for the electric connection.

**Author contributions** KBF and BBJ conceived the idea; XYC and KBF performed the simulations; YCX prepared the samples and carried out experimental measurements; YCX, JQY and DCZ implemented the optimization algorithms; SW and JBW helped on the sample fabrication. CHZ helped on the measurement setup, JC, WJP, and PHW provided a constructive discussion on the idea and the manuscript. YCX, JQY and KBF analyzed the data. YCX, JQY, KBF and WJP wrote the manuscript; KBF and BBJ supervised the project. YCX and JQY. contribute equally to this work.

**Competing financial interests** The authors declare no competing financial interests.

# References:

[1] R. Juliano Martins, E. Marinov, M. A. B. Youssef, C. Kyrou, M. Joubert, C. Colmagro, V. Gâtˊe, C. Turbil, P.-M. Coulon, D. Turover, et al., Nature Communications **13**, 5724 (2022).

[2] J. Park, B. G. Jeong, S. I. Kim, D. Lee, J. Kim, C. Shin, C. B. Lee, T. Otsuka, J. Kyoung, S. Kim, *et al.*, Nature nanotechnology **16**, 69 (2021).

[3] W. Tang, M. Z. Chen, J. Y. Dai, Y. Zeng, X. Zhao, S. Jin, Q. Cheng, and T. J. Cui, IEEE Wireless Communications **27**, 180 (2020).

[4] V. G. Ataloglou, S. Taravati, and G. V. Eleftheriades, National Science Review **10**, nwad164 (2023).

[5] N. Kanda, K. Konishi, N. Nemoto, K. Midorikawa, and M. Kuwata-Gonokami, Scientific reports **7**, 42540 (2017).

[6] M. Pan, Y. Fu, M. Zheng, H. Chen, Y. Zang, H. Duan, Q. Li, M. Qiu, and Y. Hu, Light: Science & Applications **11**, 195 (2022).

[7] F. Lan, L. Wang, H. Zeng, S. Liang, T. Song, W. Liu, P. Mazumder, Z. Yang, Y. Zhang, and D. M. Mittleman, Light: Science & Applications **12**, 191 (2023).

[8] G.-Y. Lee, J.-Y. Hong, S. Hwang, S. Moon, H. Kang, S. Jeon, H. Kim, J.-H. Jeong, and B. Lee, Nature communications **9**, 4562 (2018).

[9] Z. Li, R. Pestourie, J.-S. Park, Y.-W. Huang, S. G. Johnson, and F. Capasso, Nature communications **13**, 2409 (2022).

[10] Z. Liu, D. Wang, H. Gao, M. Li, H. Zhou, and C. Zhang, Advanced Photonics **5**, 034001(2023).

[11] H. Sarieddeen, M.-S. Alouini, and T. Y. Al-Naffouri, Proceedings of the IEEE **109**, 1628(2021).

[12] T. Nagatsuma, G. Ducournau, and C. C. Renaud, Nature Photonics **10**, 371 (2016).

[13] A. Araghi, M. Khalily, M. Safaei, A. Bagheri, V. Singh, F. Wang, and R. Tafazolli, IEEEAccess **10**, 2646 (2022).

[14] L. Dai, B. Wang, M. Wang, X. Yang, J. Tan, S. Bi, S. Xu, F. Yang, Z. Chen, M. Di Renzo, *et al.*, IEEE access **8**, 45913 (2020).

[15] T. J. Cui, M. Q. Qi, X. Wan, J. Zhao, and Q. Cheng, Light: science & applications **3**, e218 (2014).

[16] H. Yang, F. Yang, S. Xu, Y. Mao, M. Li, X. Cao, and J. Gao, IEEE Transactions on Antennas and Propagation **64**, 2246 (2016).

[17] X. Gao, W. L. Yang, H. F. Ma, Q. Cheng, X. H. Yu, and T. J. Cui, IEEE Transactions onAntennas and Propagation **66**, 6086 (2018).

[18] D. Shrekenhamer, S. Rout, A. C. Strikwerda, C. Bingham, R. D. Averitt, S. Sonkusale, and W. J. Padilla, Optics express **19**, 9968 (2011).

[19] D. Shrekenhamer, W.-C. Chen, and W. J. Padilla, Physical review letters **110**, 177403 (2013).

[20] S. Savo, D. Shrekenhamer, and W. J. Padilla, Advanced optical materials **2**, 275 (2014).

[21] C. M. Watts, D. Shrekenhamer, J. Montoya, G. Lipworth, J. Hunt, T. Sleasman, S. Krishna, D. R. Smith, and W. J. Padilla, Nature photonics **8**, 605 (2014).

[22] D. Wang, L. Zhang, Y. Gu, M. Mehmood, Y. Gong, A. Srivastava, L. Jian, T. Venkatesan, C.-W. Qiu, and M. Hong, Scientific reports **5**, 15020 (2015).

[23] M. Liu, H. Y. Hwang, H. Tao, A. C. Strikwerda, K. Fan, G. R. Keiser, A. J. Sternbach, K. G. West, S. Kittiwatanakul, J. Lu, *et al.*, Nature **487**, 345 (2012).

[24] Y. Zhang, Y. Zhao, S. Liang, B. Zhang, L. Wang, T. Zhou, W. Kou, F. Lan, H. Zeng, J. Han, *et al.*, Nanophotonics **8**, 153 (2018).

[25] J. Park, S. J. Kim, P. Landreman, and M. L. Brongersma, Advanced Optical Materials **8**, 2000745 (2020).

[26] M. Tamagnone, S. Capdevila, A. Lombardo, J. Wu, A. Centeno, A. Zurutuza, A. M. Ionescu, A. C. Ferrari, and J. R. Mosig, arXiv preprint arXiv:1806.02202 (2018).

[27] J. Wu, Z. Shen, S. Ge, B. Chen, Z. Shen, T. Wang, C. Zhang, W. Hu, K. Fan, W. Padilla, et al., Applied physics letters **116** (2020).

[28] B. Chen, X. Wang, W. Li, C. Li, Z. Wang, H. Guo, J. Wu, K. Fan, C. Zhang, Y. He, et al., Science Advances **8**, eadd1296 (2022).

[29] W. Li, B. Chen, X. Hu, H. Guo, S. Wang, J. Wu, K. Fan, C. Zhang, H. Wang, B. Jin, et al., Science Advances **9**, eadi7565 (2023).

[30] S. Molesky, Z. Lin, A. Y. Piggott, W. Jin, J. Vuckovićc, and A. W. Rodriguez, Nature Photonics **12**, 659 (2018).

[31] Z. Li, R. Pestourie, Z. Lin, S. G. Johnson, and F. Capasso, ACS Photonics **9**, 2178 (2022).

[32] Z. Liu, D. Zhu, S. P. Rodrigues, K.-T. Lee, and W. Cai, Nano letters **18**, 6570 (2018).

[33] P. R. Wiecha, A. Arbouet, C. Girard, and O. L. Muskens, Photonics Research **9**, B182 (2021).

[34] R. Zhu, T. Qiu, J. Wang, S. Sui, C. Hao, T. Liu, Y. Li, M. Feng, A. Zhang, C.-W. Qiu, et al., Nature communications **12**, 2974 (2021).

[35] C.-H. Lin, Y.-S. Chen, J.-T. Lin, H. C. Wu, H.-T. Kuo, C.-F. Lin, P. Chen, and P. C. Wu, Nano letters **21**, 4981 (2021).

[36] P. Thureja, G. K. Shirmanesh, K. T. Fountaine, R. Sokhoyan, M. Grajower, and H. A. Atwater, ACS nano **14**, 15042 (2020).

[37] J.-Q. Yang, Y. Xu, J.-L. Shen, K. Fan, D.-C. Zhan, and Y. Yang, in Proceedings of the 29th ACM SIGKDD Conference on Knowledge Discovery and Data Mining (2023) pp. 2930–2940.

[38] O. Khatib, S. Ren, J. Malof, and W. J. Padilla, Advanced Functional Materials **31**, 2101748 (2021).

[39] Z. Liu, D. Zhu, L. Raju, and W. Cai, Advanced Science **8**, 2002923 (2021).

[40] C. Liu, Q. Ma, Z. J. Luo, Q. R. Hong, Q. Xiao, H. C. Zhang, L. Miao, W. M. Yu, Q. Cheng, L. Li, *et al.*, Nature Electronics **5**, 113 (2022).

[41] L. Gao, X. Li, D. Liu, L. Wang, and Z. Yu, Advanced Materials **31**, 1905467 (2019).

[42] S. An, B. Zheng, H. Tang, M. Y. Shalaginov, L. Zhou, H. Li, M. Kang, K. A. Richardson, T. Gu, J. Hu, *et al.*, Advanced Optical Materials **9**, 2001433 (2021).

[43] J. Chen, C. Qian, J. Zhang, Y. Jia, and H. Chen, Nature Communications **14**, 4872 (2023).

[44] Z. A. Kudyshev, A. V. Kildishev, V. M. Shalaev, and A. Boltasseva, Applied Physics Reviews **7** (2020).

[45] Y. Deng, S. Ren, K. Fan, J. M. Malof, and W. J. Padilla, Optics Express **29**, 7526 (2021).

[46] J.-Q. Yang, Y. Xu, K. Fan, J. Wu, C. Zhang, D.-C. Zhan, B.-B. Jin, and W. J. Padilla, ACS Photonics **10**, 1001 (2023).

[47] G. E. Karniadakis, I. G. Kevrekidis, L. Lu, P. Perdikaris, S. Wang, and L. Yang, Nature Reviews Physics **3**, 422 (2021).

[48] S. Cuomo, V. S. Di Cola, F. Giampaolo, G. Rozza, M. Raissi, and F. Piccialli, Journal of Scientific Computing **92**, 88 (2022).

[49] S. Cai, Z. Wang, S. Wang, P. Perdikaris, and G. E. Karniadakis, Journal of Heat Transfer **143**, 060801 (2021).

[50] S. Cai, Z. Mao, Z. Wang, M. Yin, and G. E. Karniadakis, Acta Mechanica Sinica **37**, 1727 (2021).

[51] J. Zhou, R. Li, and T. Luo, npj Computational Materials **9**, 212 (2023).

[52] S. Sarkar, A. Ji, Z. Jermain, R. Lipton, M. Brongersma, K. Dayal, and H. Y. Noh, Advanced Photonics Research **4**, 2300158 (2023).

[53] Y. Chen, L. Lu, G. E. Karniadakis, and L. Dal Negro, Optics express **28**, 11618 (2020).

[54] Y. Tang, J. Fan, X. Li, J. Ma, M. Qi, C. Yu, and W. Gao, Nature Computational Science **2**, 169 (2022).

[55] J. Jiang and J. A. Fan, Nano letters **19**, 5366 (2019).

[56] Z. Fang and J. Zhan, Ieee Access **8**, 24506 (2019).

[57] L. Zhang, X. Q. Chen, S. Liu, et al., Nature Communications **9**, 4334 (2018).

[58] B. E. Saleh and M. C. Teich, *Fundamentals of photonics* (john Wiley & sons, 2019).

[59] G. Isi´c, B. Vasi´c, D. C. Zografopoulos, R. Beccherelli, and R. Gaji´c, Physical Review Applied **3**, 064007 (2015).

[60] J. Park, J.-H. Kang, S. J. Kim, X. Liu, and M. L. Brongersma, Nano letters **17**, 407 (2017).

[61] N. Yu, P. Genevet, M. A. Kats, F. Aieta, J.-P. Tetienne, F. Capasso, and Z. Gaburro, science **334**, 333 (2011).

[62] L. Huang, X. Chen, H. Muhlenbernd, G. Li, B. Bai, Q. Tan, G. Jin, T. Zentgraf, and S. Zhang, Nano letters **12**, 5750 (2012).

[63] Y. Xie, W. Wang, H. Chen, A. Konneker, B.-I. Popa, and S. A. Cummer, Nature communications **5**, 5553 (2014).

[64] R. Audhkhasi, B. Zhao, S. Fan, Z. Yu, and M. L. Povinelli, Optics Express **30**, 9463 (2022).

[65] L. Wang, S. Ge, W. Hu, M. Nakajima, and Y. Lu, Optics Express **25**, 23873 (2017).

**FIGURES**

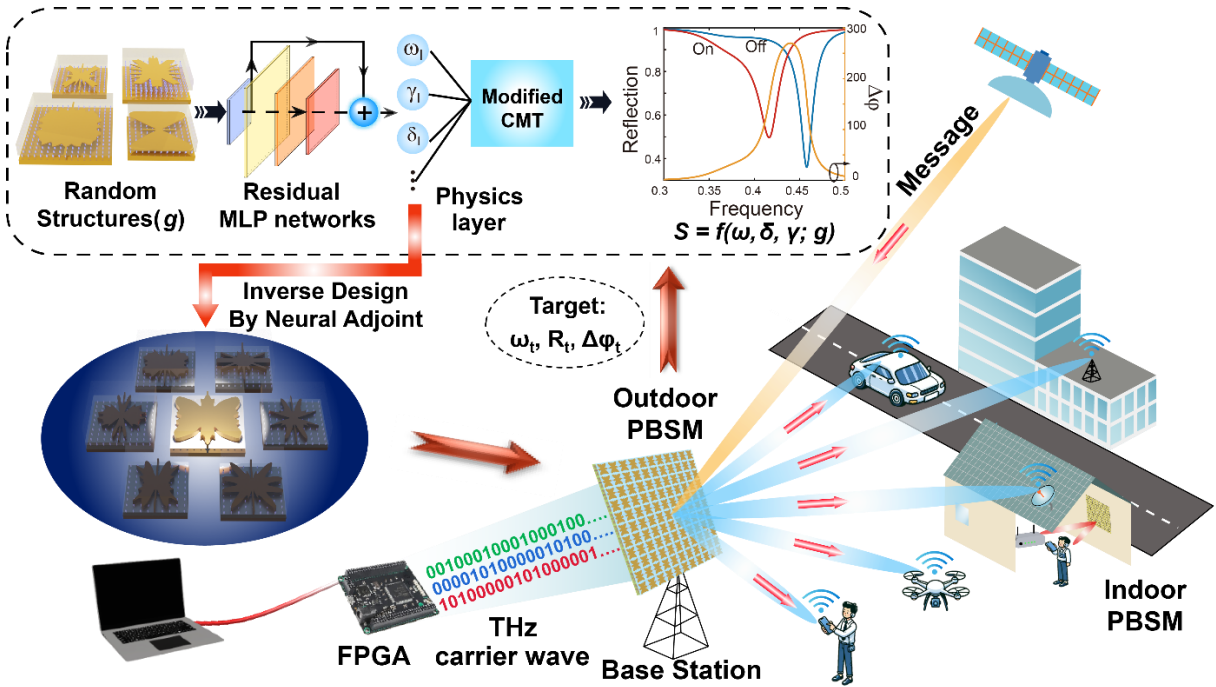

FIG. 1. Illustration of a multi-bit programmable beam-steering metasurface (PBSM). The design targets of the reflection amplitude $R_t$ at the operational frequency of $\omega_t$, and the maximum phase change $\Delta\varphi_t$ are achieved by the proposed physics-informed inverse design (PIID) method. In the inverse design flow, a modified coupled-mode theory (MCMT) with the Fabry-Pérot effect is integrated into a residual multilayer perceptron (ResMLP) network to learn the physics between the geometric and the MCMT parameters. The inverse designed programmable metasurface can steer the incident terahertz beams for telecommunications with multiple receivers.

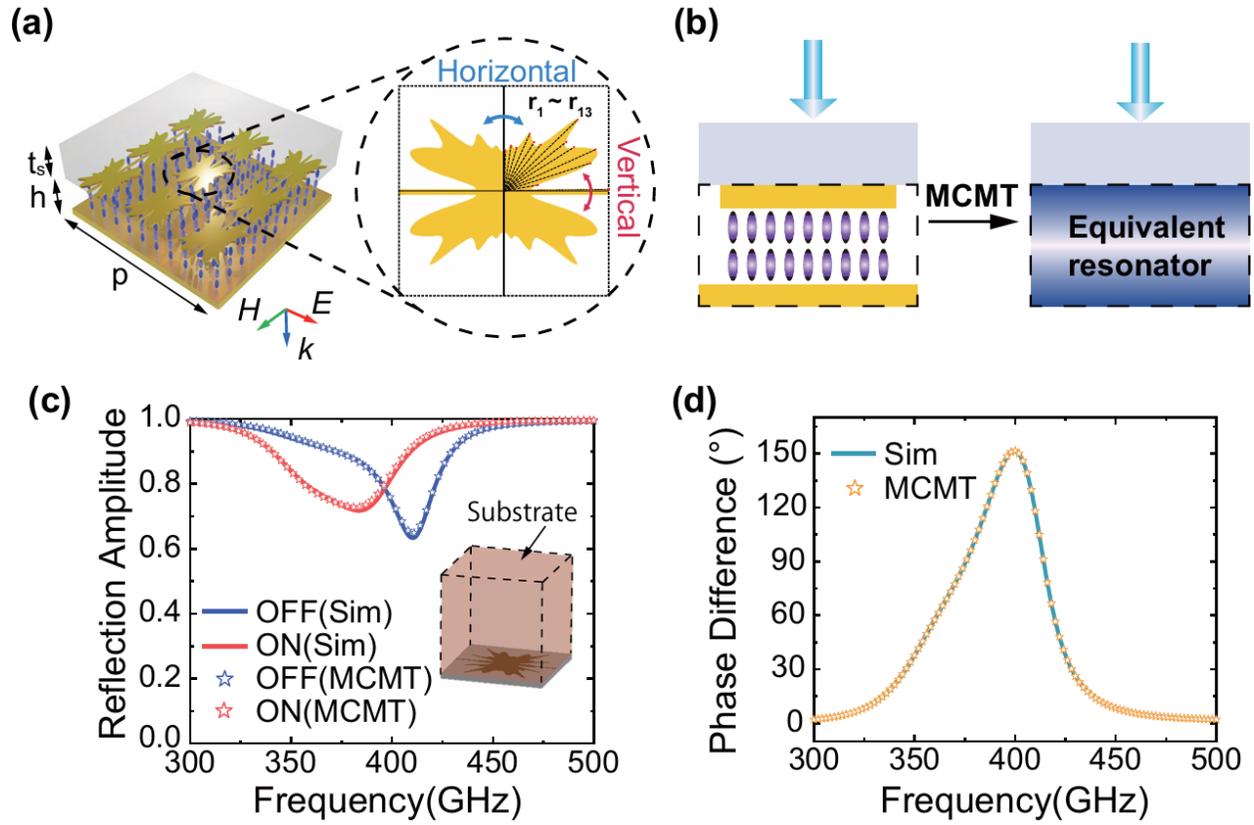

FIG. 2. Theoretical analysis for metasurface with substrate supported according to proposed MCMT. (a) Details of a freeform structure generated using the B-spline algorithm. (b) Schematic diagram of MCMT analysis. Reflection Amplitude spectrum (c) and Phase Difference spectrum(d) of the metasurface with substrate precisely calculated by MCMT (comparing to simulations)

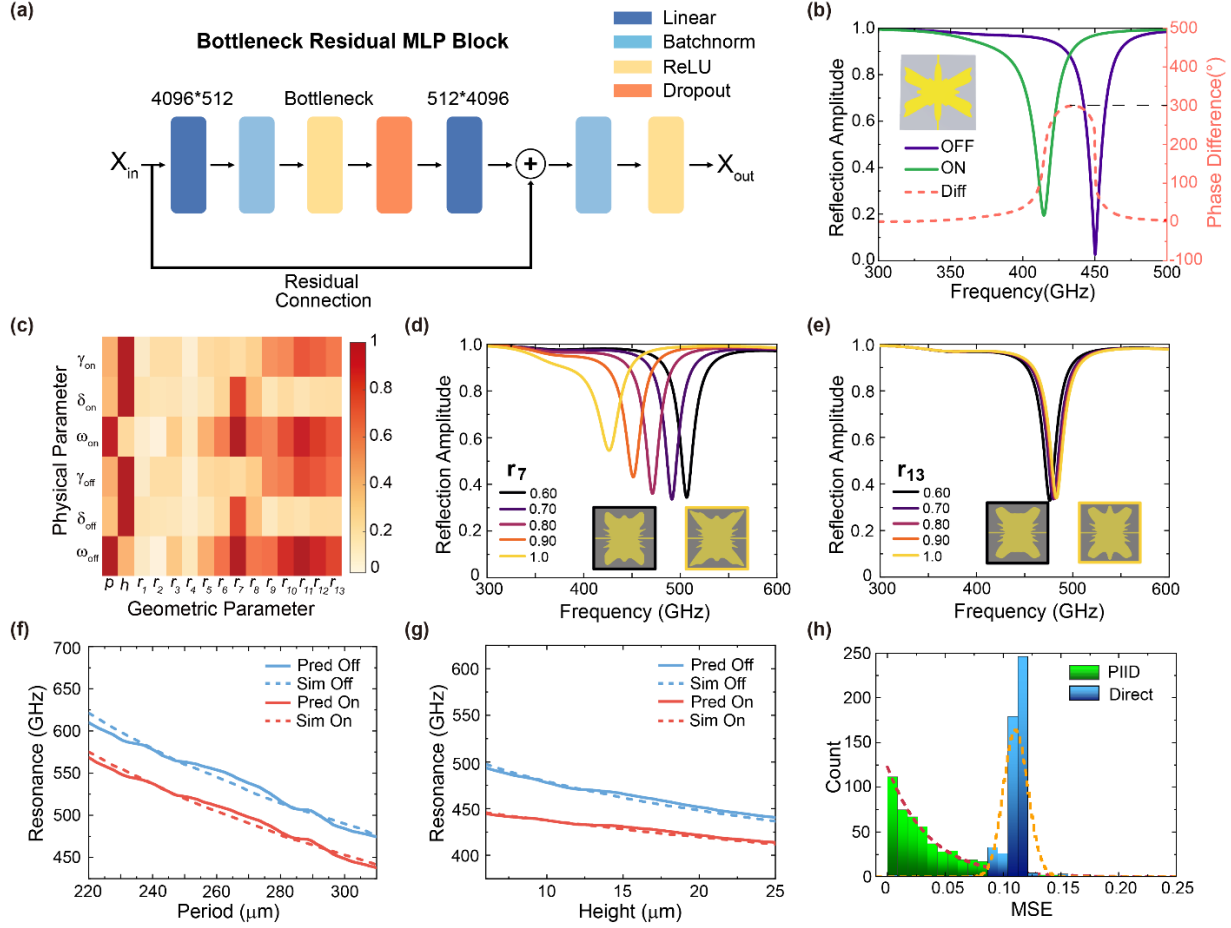

FIG. 3. Inversely designing the programmable metasurface. (a) Detailed illustration of the structure of Bottleneck Residue MLP Block. (b) An example of an inversely designed tunable LC-based metasurface with a phase change near 300° before and after modulation. (c) The relative influence of geometric parameters on mode parameters, with the LC layer thickness $h$ and control parameter $x_7$ exhibiting the greatest impact. By varying the values of $x_7$ and $x_{13}$ respectively (d and e), a stronger dependence of the spectrum on the parameter $x_7$ is observed according to simulations. (f) The correlation between the resonant frequency and the periodicity predicted by the ResMLP and verified by simulations. (g) The correlation between the resonant frequency and LC layer thickness predicted by the ResMLP and verified by simulations. (h) Comparison of MSE distribution for inversely designed metasurfaces by ResMLP network with and without the physics layers.

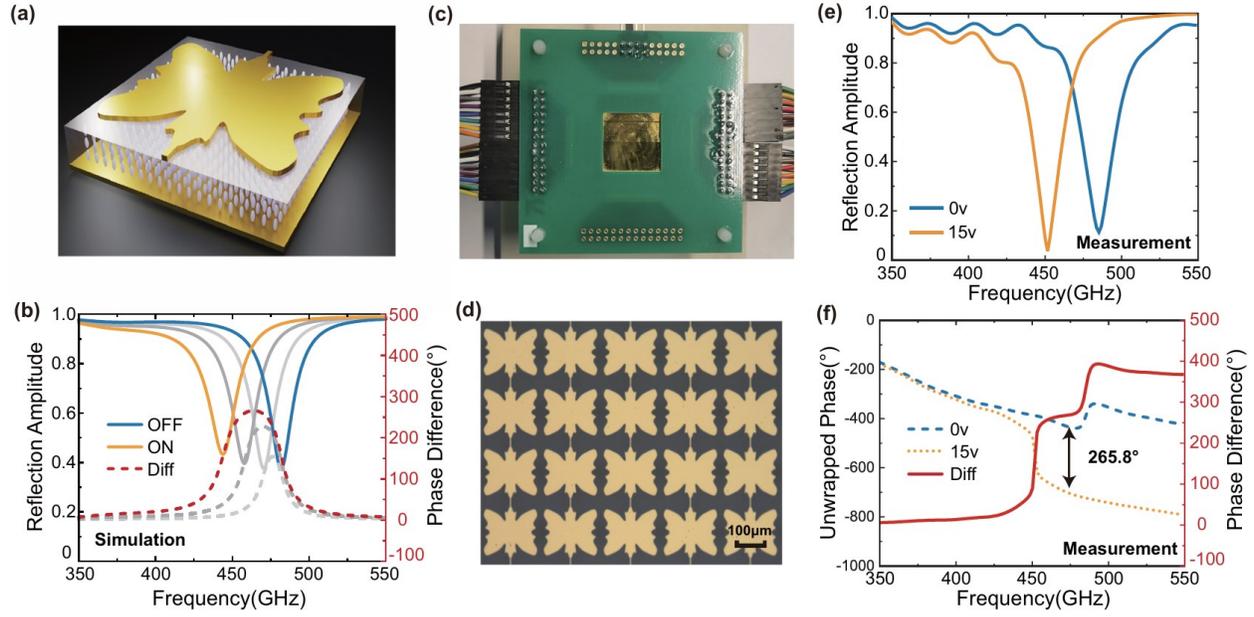

FIG. 4. Demonstration of an inversely designed terahertz PBSM. (a) Schematic of the designed unit cell structure. (b) Simulated reflection spectra in 'OFF' and 'ON' states and phase difference between two states. The grey lines indicate the intermediate states with the applied voltage between the minimum and maximum values. (c) Test photograph of the half-modulated(/0..01..1/) metasurface. (d) Microscopic image of the fabricated metasurface. Measured amplitude (e) and phase (f) spectra with the bias voltage at 0v (blue) and 15v (yellow). The phase change between the two states is also calculated and shown as the solid red curve.

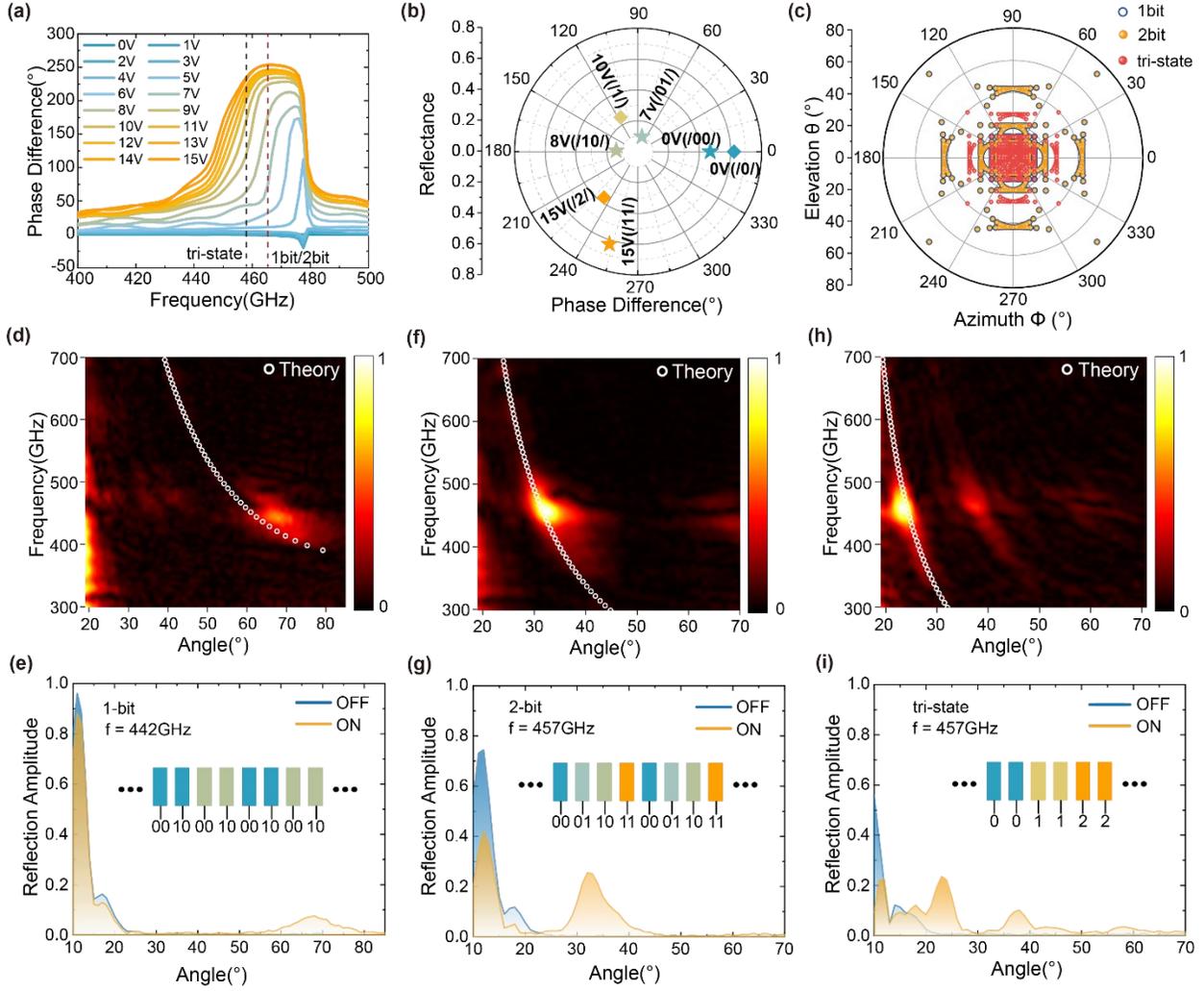

FIG. 5. Measurements of the beam steering for the fabricated PBSM. (a) The phase difference spectra of the metasurface with varying bias (1V interval). The dashed red and black curves denote the theoretical operating frequencies of the metasurface under 1-bit/2-bit and tri-state coding schemes, respectively. (b) The applied voltage in experiments and corresponding coding notation for the 1-bit/2-bit and tri-state methods. (c) Calculation of deflection angle coverage in the upper half-plane using a row-column controlled PBSM based on the designed metasurface. Experiment results of the beam deflection when the device is encoded with various sequences. (d) and (e) 1-bit coding with the sequence of /00 10/; (f) and (g) 2-bit coding with the sequence of /00 01 10 11/; (h) and (I) tri-state coding with the sequence of /00 11 22/. The top row shows the measured beam reflection spectra as a function of the receiver angle for the three coding schemes. The bottom row shows the beam steering performance at 442 GHz, 457 GHz, and 457 GHz, respectively.